\documentclass[preprint,showpacs,preprintnumbers,amsmath,amssymb,endfloats*]{revtex4}
\usepackage{graphicx}

\begin{document}
\def \ee {\varepsilon}
\thispagestyle{empty}
\title{
Van der Waals interaction between a microparticle and a single-wall
carbon nanotube}

\author{E.~V.~Blagov,${}^{1}$
G.~L.~Klimchitskaya,${}^{2}$
and V.~M.~Mostepanenko,${}^{1}$
}

\affiliation{
${}^{1}$Noncommercial Partnership ``Scientific Instruments'',
Tverskaya St. 11, Moscow, 103905, Russia\\
${}^{2}$North-West Technical University, Millionnaya St. 5, St.Petersburg,
191065, Russia
}

\begin{abstract}
The Lifshitz-type formulas describing the free energy and the force
of the van der Waals interaction between an atom (molecule) and
a single-wall carbon nanotube are obtained. The single-wall
nanotube is considered as a cylindrical sheet carrying a two-dimensional
free electron gas with appropriate boundary conditions on the
electromagnetic field. The obtained formulas are used to
calculate the van der Waals free energy and force between a
hydrogen atom (molecule) and single-wall carbon nanotubes of different
radia. Comparison studies of the van der Waals interaction
of hydrogen atoms with single- and multi-wall carbon nanotubes
show that depending on atom-nanotube separation distance
the idealization of graphite dielectric permittivity is already
applicable to nanotubes with only two or three walls.
\end{abstract}
\maketitle

\section{Introduction}

Van der Waals forces play a dominant role in the interaction of
atoms and molecules with carbon nanostructures at separations
larger than about  one nanometer. These forces are of fluctuation
nature and are described in second order perturbation theory as
a dipole-dipole interaction. Theoretical description of the van
der Waals forces as a quantum phenomenon was pioneered by London.
The general theory of the van der Waals forces in the framework of
quantum statistical physics was developed by Lifshitz \cite{1,2}
using the concept of a frequency-dependent dielectric
permittivity. Lifshitz's theory describes both the 
interaction between two semispaces
or plates of finite thickness separated by a gap and between a
microparticle (an atom or a molecule) and a flat surface of a
macrobody. In the latter case it is commonly referred to as
the Casimir-Polder force \cite{3}.

An understanding of the underlying mechanisms of
microparticle---nanostructure interaction is of importance for
many applications including the problem of hydrogen storage in
carbon nanostructures \cite{4}. However, the application of the
Lifshitz theory to the latter case encounters serious problems
because the boundary surface of carbon nanotubes is not flat
and single-wall nanotubes are not characterized by the
macroscopic concept of a frequency-dependent dielectric
permittivity. Because of this, most theoretical work on the
van der Waals interaction in layered structures and between
hydrogen atoms (molecules) and a graphite sheet or carbon nanotubes
was done using the phenomenological density functional theory
(see, e.g., Refs.~\cite{5,6,7,8,9,10,11}). Some other approximate
approaches were also used, e.g., the nonrelativistic perturbation
theory for degenerate levels of a two-level atomic system
\cite{12,13}. It is known, however, that in some cases the
approximate and phenomenological approaches do not provide a
precise description of the van der Waals interaction \cite{14,14a}.
This is true \cite{15}, for instance, when one uses the density
functional theory (especially with linear-density approximation).

Recently, the scope of the Lifshitz theory of the van der Waals
forces was much widened by successful application to the
interpretation of precision measurements of the Casimir force
\cite{16,17,18,19,20,21,22} (i.e., the van der Waals force at
larger separations where the relativistic retardation becomes
essential) and to atom-wall interaction in connection with
Bose-Einstein condensation \cite{23,24}. What is more, in
Refs.~\cite{25,26} the Lifshitz theory was extended for the
case of an atom (molecule) interacting with a plane surface of
a uniaxial crystal or with a multi-wall carbon nanotube. The
generalization to nanotubes was achieved by using the proximity
force theorem \cite{27} and the idealization of graphite
dielectric permittivities which is applicable to multi-wall
nanotubes with sufficiently large number of walls. In
Ref.~\cite{28} the Lifshitz-type formulas were obtained for the
van der Waals interaction between a single layer of graphite
(hereafter, graphene) and a material plate, graphene and an
atom or a molecule, and between a single-wall carbon nanotube
and a plate. To achieve this goal, graphene was considered in
terms of a two-dimensional free electron gas \cite{29}.
The reflection properties of electromagnetic oscillations on
graphene were described by the specific boundary conditions
imposed on the infinitely thin positively charged plasma sheet,
carrying a continuous fluid with some mass and negative charge
density \cite{30}.

In the present paper we use the same model of graphene as in
Ref.~\cite{28} and obtain Lifshitz-type formulas for the
van der Waals interaction between a microparticle (an atom or
a molecule) and a single-wall carbon nanotube. Both the free
energy and interaction force at arbitrary temperature are
considered. As an example, we have calculated the van der
Waals free energy and force  as functions of separation
in the configuration of a hydrogen
atom and a molecule in close proximity to a single-wall carbon
nanotube. Specifically, the values of 
the van der Waals coefficient
are determined at different separations. Comparison
studies of the van der Waals interaction of a microparticle
with multi- and single-wall carbon nanotubes are performed.
If we consider the van der Waals interaction of
an atom or a molecule with
a nanotube having only two or three walls, the idealization
of the dielectric permittivity of graphite is demonstrated
to be already applicable if separation distance between a 
microparticle and a nanotube is sufficiently large.

The paper is organized as follows. In Sec.~II we derive the
Lifshitz-type formulas for the van der Waals free energy
and force acting between a microparticle
and a single-wall carbon nanotube. Sec.~III is devoted to the
numerical computations of the van der Waals interaction between
a hydrogen atom or a molecule and a single-wall nanotube. In Sec.~IV
the comparison between the cases of single- and multi-wall
nanotubes is performed. Sec.~V contains our conclusions and
discussion.

\section{Lifshitz-type formulas for the interaction between
\protect{\\}
microparticle and single-wall carbon nanotube}

We begin with the van der Waals interaction of a graphene
occupying the $xy$-plane, $z=0$, or of a graphite plate of
thickness $d$, with a semispace made of isotropic material.
The separation distance between the graphene or the boundary plane of
a graphite plate (labeled by the upper index 1) and the boundary
plane of a semispace (labeled by 2) is $a$. As was shown in
Refs.~\cite{25,26,28}, in both cases the free energy of the
van der Waals interaction per unit area at temperature $T$ in
thermal equilibrium is given by the Lifshitz-type formula with
the properly defined reflection coefficients
$r_{\rm TM,TE}^{(1,2)}$:
\begin{eqnarray}
&&
{\cal F}(a,T)=\frac{k_BT}{2\pi}\sum\limits_{l=0}^{\infty}
\left(1-\frac{1}{2}\delta_{l0}\right)\int_{0}^{\infty}
k_{\bot}\,dk_{\bot} \nonumber \\
&&
\phantom{aaa}\times\left\{
\ln\left[1-r_{\rm TM}^{(1)}(\xi_l,k_{\bot})
r_{\rm TM}^{(2)}(\xi_l,k_{\bot})e^{-2aq_l}\right]\right.
\label{eq1} \\
&&
\phantom{aaaa}+\left.
\ln\left[1-r_{\rm TE}^{(1)}(\xi_l,k_{\bot})
r_{\rm TE}^{(2)}(\xi_l,k_{\bot})e^{-2aq_l}\right]\right\}.
\nonumber
\end{eqnarray}
\noindent
Here, $k_B$ is the Boltzmann constant, $\xi_l=2\pi k_BTl/\hbar$
with $l=1,\,2,\,3,\,\ldots$ are the Matsubara frequencies along
the imaginary frequency axis ($\omega=i\xi$), $k_{\bot}$ is the
magnitude of the wave vector component perpendicular to the
$z$-axis, transverse electric (TE) and transverse magnetic (TM)
modes are the two independent polarizations of electromagnetic
field, and $q_l=\left(k_{\bot}^2+\xi_l^2/c^2\right)^{1/2}$.

From Eq.~(\ref{eq1}), the van derWaals force acting between a
graphene or a graphite plate of thickness $d$ and a semispace
made of isotropic material is:
\begin{eqnarray}
&&
F(a,T)=-\frac{\partial{\cal F}(a,T)}{\partial a}=
-\frac{k_BT}{\pi}\sum\limits_{l=0}^{\infty}
\left(1-\frac{1}{2}\delta_{l0}\right)\int_{0}^{\infty}
k_{\bot}\,dk_{\bot}\,q_l \nonumber \\
&&
\phantom{aaa}\times\left[
\frac{r_{\rm TM}^{(1)}(\xi_l,k_{\bot})
r_{\rm TM}^{(2)}(\xi_l,k_{\bot})}{e^{2aq_l}-
r_{\rm TM}^{(1)}(\xi_l,k_{\bot})
r_{\rm TM}^{(2)}(\xi_l,k_{\bot})}\right.
\label{eq2} \\
&&
\phantom{aaaa}+\left.
\frac{r_{\rm TE}^{(1)}(\xi_l,k_{\bot})
r_{\rm TE}^{(2)}(\xi_l,k_{\bot})}{e^{2aq_l}-
r_{\rm TE}^{(1)}(\xi_l,k_{\bot})
r_{\rm TE}^{(2)}(\xi_l,k_{\bot})}\right].
\nonumber
\end{eqnarray}

Now we specify the reflection coefficients. For a semispace made
of isotropic material (labeled by the upper index 2) they are
commonly known \cite{2,17}
\begin{equation}
r_{\rm TM}^{(2)}(\xi_l,k_{\bot})=\frac{\varepsilon(i\xi_l)q_l-
k_l}{\varepsilon(i\xi_l)q_l+k_l},
\quad
r_{\rm TE}^{(2)}(\xi_l,k_{\bot})=\frac{k_l-q_l}{k_l+q_l},
\label{eq3}
\end{equation}
\noindent
where
\begin{equation}
k_l=\sqrt{k_{\bot}^2+\varepsilon(i\xi_l)\frac{\xi_l^2}{c^2}},
\label{eq4}
\end{equation}
\noindent
and $\varepsilon(\omega)$ is the dielectric permittivity of a
semispace material.

If the first body is a graphite plate of thickness $d$, the
reflection coefficients were found in Ref.~\cite{25}:
\begin{eqnarray}
&&
r_{\rm TM}^{(1)}(\xi_l,k_{\bot})=r_{{\rm TM},d}^{(1)}(\xi_l,k_{\bot})
=\frac{\varepsilon_x(i\xi_l)\varepsilon_z(i\xi_l)q_l^2-
k_{zl}^2}{\varepsilon_x(i\xi_l)\varepsilon_z(i\xi_l)q_l^2+k_{zl}^2+
2q_lk_{zl}\sqrt{\varepsilon_x(i\xi_l)\varepsilon_z(i\xi_l)}
\coth(k_{zl}d)},
\nonumber \\
&&
r_{\rm TE}^{(1)}(\xi_l,k_{\bot})=r_{{\rm TE},d}^{(1)}(\xi_l,k_{\bot})
=\frac{k_{xl}^2-q_l^2}{k_{xl}^2+q_l^2+2q_lk_{xl}\coth(k_{xl}d)}.
\label{eq5}
\end{eqnarray}
\noindent
Here, $\varepsilon_x(\omega)=\varepsilon_y(\omega)$ and
$\varepsilon_z(\omega)$ are the dielectric permittivities of
graphite
in the $x,\,y$ and $z$ directions, respectively, and
\begin{equation}
k_{xl}=\sqrt{k_{\bot}^2+\varepsilon_x(i\xi_l)\frac{\xi_l^2}{c^2}},
\quad
k_{zl}=\sqrt{k_{\bot}^2+\varepsilon_z(i\xi_l)\frac{\xi_l^2}{c^2}}.
\label{eq6}
\end{equation}

If the first body is graphene, the idealization of the
frequency-dependent dielectric permittivity cannot be used.
In this case the reflection coefficients can be found
\cite{28,30,31} by modeling graphene as a two-dimensional plasma
sheet carrying a negatively charged fluid of $\pi$-electrons.
For the hexagonal
structure of carbon layers there is one $\pi$-electron per
atom \cite{32} resulting in two $\pi$-electrons per one hexagonal cell.
This leads to the following values for the density of
$\pi$-electrons and the wave number of the sheet:
\begin{equation}
n=\frac{4}{3\sqrt{3}l^2},
\quad
K=2\pi\frac{ne^2}{mc^2}=6.75\times10^5\,\mbox{m}^{-1},
\label{eq7}
\end{equation}
\noindent
where $l=1.421\,${\AA} is the side length of a hexagon, $e$ and $m$
are the electron charge and mass, respectively.

Now we are solving the Maxwell equations with the following
matching conditions on the tangential and normal
components of the fields:
\begin{eqnarray}
&&
\mbox{\boldmath$E$}_{t,2}-\mbox{\boldmath$E$}_{t,1}=0, \quad
E_{z,2}-E_{z,1}=2K\frac{c^2}{\omega^2}
\nabla_{\! t}\cdot\mbox{\boldmath$E$}_t,
\label{eq8} \\
&&
B_{z,2}-B_{z,1}=0, \quad
\mbox{\boldmath$B$}_{t,2}-\mbox{\boldmath$B$}_{t,1}=
-2iK\frac{c}{\omega}\mbox{\boldmath$j$}
\times\mbox{\boldmath$E$}_{t}.
\nonumber
\end{eqnarray}
\noindent
Here $\mbox{\boldmath$j$}=(0,\,0,\,1)$ is the unit vector pointing
in $z$-direction, and all fields depend on time through a common
factor exp$(-i\omega t)$.
As a result, we arrive to one-dimensional scattering
problem in the $z$-direction \cite{17}. The solution of this
problem leads to the following reflection coefficients on a
graphene sheet \cite{28,30,31}:
\begin{eqnarray}
&&
r_{\rm TM}^{(1)}(\xi_l,k_{\bot})=r_{{\rm TM},g}^{(1)}(\xi_l,k_{\bot})
=\frac{c^2q_lK}{c^2q_lK+\xi_l^2},
\nonumber \\
&&
r_{\rm TE}^{(1)}(\xi_l,k_{\bot})=r_{{\rm TE},g}^{(1)}(\xi_l,k_{\bot})
=\frac{K}{K+q_l}.
\label{eq9}
\end{eqnarray}

Now we use Eqs.~(\ref{eq1}) and (\ref{eq2}) with the reflection
coefficients (\ref{eq3}) and (\ref{eq9}) to obtain the free energy
of van der Waals interaction and force acting between an atom
(molecule) and a single-wall carbon nanotube. For this purpose
let us consider an infinite space filled with an isotropic
substance having a dielectric permittivity $\varepsilon(\omega)$
containing an empty cylindrical cavity of radius $R+a$. We
introduce the single-wall carbon nanotube of radius $R$ inside
this cavity (see Fig.~1). In so doing the nanotube is modeled by
a cylindrical graphene sheet. Then there is a gap of thickness $a$
between the nanotube and the boundary surface of the cylindrical
cavity. Each element of the nanotube experiences an attractive
van der Waals force on the source side of this boundary surface.
By using the proximity force approximation (see Ref.~\cite{33}
in the case of ideal metals), the free energy of the van der Waals
interaction between these two co-axial cylinders can be
approximately represented in the form
\begin{equation}
{\cal F}^{c,c}(a,T)=2\pi L\sqrt{R(R+a)}{\cal F}(a,T).
\label{eq10}
\end{equation}
\noindent
Here ${\cal F}(a,T)$ is the free energy per unit area in the
configuration of a graphene interacting with a plane boundary of
an isotropic semispace determined in Eq.~(\ref{eq1}), and $L$
is the length of the nanotube, which is supposed to be much
larger than $R$.

The accuracy of Eq.~(\ref{eq10}) is rather high. Recently the
accuracy of the proximity force approximation was investigated
on the basis of first principles by comparing the approximate
results with the exact ones for the configurations of a cylinder or
a sphere above plate. It was shown \cite{34,35,36,37} that
corrections to the results obtained using the proximity force
approximation are less than $0.5a/R$ where $R$ is a cylinder or
sphere radius. For the configuration of two co-axial cylinders
the accuracy of the proximity force approximation is even much
higher. For example, within the separation region
$0<a<R/2$, the results calculated using Eq.~(\ref{eq10})
coincide with the exact ones up to 1\% for cylinders
made of ideal metal \cite{33,38} (for real materials the
accuracy may be different for only a fraction of a percent).

To come to the case of an atom (or molecule) near a nanotube, we
suppose that the isotropic substance filling the infinite space
is rarefied with some small number $N$ of atoms (molecules)
per unit volume. We expand the quantity ${\cal F}^{c,c}(a,T)$
on the left-hand side of Eq.~(\ref{eq10}) in powers of $N$
and using the additivity of the first-order term arrive at:
\begin{equation}
{\cal F}^{c,c}(a,T)=N\int_{a}^{\infty}{\cal F}^{a,c}(z,T)
2\pi(R+z)Ldz+\mbox{O}(N^2).
\label{eq11}
\end{equation}
\noindent
Here, ${\cal F}^{a,c}(z,T)$ is (yet not found) the free energy
of the van der Waals interaction of a single atom (molecule)
belonging to an isotropic substance with a single-wall carbon
nanotube ($z$ is measured from the external surface of
nanotube in the direction perpendicular to it).

Differentiation of both (negative) sides of
Eq.~(\ref{eq11})
with respect to $a$ leads to:
\begin{equation}
-\frac{\partial{\cal F}^{c,c}(a,T)}{\partial a}=
2\pi(R+a)LN{\cal F}^{a,c}(a,T)+\mbox{O}(N^2).
\label{eq12}
\end{equation}

This negative derivative can be found also by the differentiation
of both sides of Eq.~(\ref{eq10}):
\begin{eqnarray}
&&
-\frac{\partial{\cal F}^{c,c}(a,T)}{\partial a}=
2\pi L\sqrt{R(R+a)}
\label{eq13}\\
&&\phantom{aaaaa}
\times\left[-\frac{1}{2(R+a)}{\cal F}(a,T)+F(a,T)\right],
\nonumber
\end{eqnarray}
\noindent
where the force $F(a,T)$ acting between a graphene and a semispace
was defined in Eq.~(\ref{eq2}) with the reflection coefficients
(\ref{eq3}) and (\ref{eq9}).

We expand the dielectric permittivity of a rarefied substance in
powers of $N$ \cite{39}
\begin{equation}
\varepsilon(i\xi_l)=1+4\pi\alpha(i\xi_l)N+\mbox{O}(N^2),
\label{eq14}
\end{equation}
\noindent
where $\alpha(\omega)$ is the dynamic polarizability of an atom or
a molecule of this substance. Using Eq.~(\ref{eq14}), we expand
also the reflection coefficients $r_{\rm TM,TE}^{(2)}(\xi_l,k_{\bot})$
in Eq.~(\ref{eq3}) in powers of $N$ and substitute the obtained
expressions in Eqs.~(\ref{eq1}) and (\ref{eq2}). Thereafter, with
the help of Eqs.~(\ref{eq13}) and (\ref{eq12}) in the limit
$N\to 0$, the desired expression for the van der Waals free energy
in the configuration of a microparticle interacting with a
single-wall carbon nanotube is obtained (see Ref.~\cite{25}
devoted to multi-wall nanotubes for the details of perturbation
expansion and limiting procedure):
\begin{eqnarray}
&&
{\cal F}^{a,c}(a,T)=-k_BT\sqrt{\frac{R}{R+a}}
\sum\limits_{l=0}^{\infty}\left(1-\frac{1}{2}\delta_{l0}\right)
\alpha(i\xi_l)
\nonumber \\
&&
\phantom{aaa}
\times\int_{0}^{\infty}k_{\bot}\,dk_{\bot}\,e^{-2aq_l}
\left[q_l-\frac{1}{4(R+a)}\right]
\label{eq15} \\
&&
\phantom{aaaaa}
\times\left\{2r_{{\rm TM},g}^{(1)}(\xi_l,k_{\bot})+
\frac{\xi_l^2}{q_l^2c^2}\left[r_{{\rm TE},g}^{(1)}(\xi_l,k_{\bot})
-r_{{\rm TM},g}^{(1)}(\xi_l,k_{\bot})\right]\right\}.
\nonumber
\end{eqnarray}
\noindent
In the limiting case $R\to\infty$, Eq.~(\ref{eq15}) coincides with
a known result for the free energy of a microparticle near a plane
surface of graphene \cite{28}. If, instead of graphene, the
cylindrical graphite shell of thickness $d$ is considered,
Eq.~(\ref{eq15}) describes \cite{25} the interaction of a
microparticle with such a shell if the reflection coefficients
$r_{{\rm TM,TE},g}^{(1)}(\xi_l,k_{\bot})$ are replaced with
$r_{{\rm TM,TE},d}^{(1)}(\xi_l,k_{\bot})$ defined
in Eq.~(\ref{eq5}).

The van der Waals force acting between a microparticle and a
single-wall carbon nanotube is obtained as a minus derivative of
Eq.~(\ref{eq15}) with respect to separation distance
\begin{eqnarray}
&&
{F}^{a,c}(a,T)=-k_BT\sqrt{\frac{R}{R+a}}
\sum\limits_{l=0}^{\infty}\left(1-\frac{1}{2}\delta_{l0}\right)
\alpha(i\xi_l)
\nonumber \\
&&
\phantom{aaa}
\times\int_{0}^{\infty}k_{\bot}\,dk_{\bot}\,e^{-2aq_l}
\left[2q_l^2-\frac{3}{8(R+a)^2}\right]
\label{eq16} \\
&&
\phantom{aaaaa}
\times\left\{2r_{{\rm TM},g}^{(1)}(\xi_l,k_{\bot})+
\frac{\xi_l^2}{q_l^2c^2}\left[r_{{\rm TE},g}^{(1)}(\xi_l,k_{\bot})
-r_{{\rm TM},g}^{(1)}(\xi_l,k_{\bot})\right]\right\}.
\nonumber
\end{eqnarray}
\noindent
By replacing the reflection coefficients
$r_{{\rm TM,TE},g}^{(1)}(\xi_l,k_{\bot})$ with
$r_{{\rm TM,TE},d}^{(1)}(\xi_l,k_{\bot})$ we return to the
expression for the force between a microparticle and a graphite
cylindrical shell of thickness $d$ \cite{25}.

\section{Calculation of the van der waals interaction between
hydrogen atom or molecule and single-wall carbon
\protect{\\} nanotube}

For the purpose of numerical computations it is convenient
to introduce the nondimensional variables under the integrals
\begin{equation}
y=2aq_l,
\qquad
\zeta_l=\frac{2a\xi_l}{c}\equiv\frac{\xi_l}{\omega_c}
\label{eq17}
\end{equation}
\noindent
and present the free energy (\ref{eq15}) and force (\ref{eq16})
in the form
\begin{equation}
{\cal F}^{a,c}(a,T)=-\frac{C_3(a,T)}{a^3},
\quad
{F}^{a,c}(a,T)=-\frac{C_F(a,T)}{a^4},
\label{eq18}
\end{equation}
\noindent
where
\begin{eqnarray}
&&
C_3(a,T)=\frac{k_BT}{8}\sqrt{\frac{R}{R+a}}\left\{
\frac{4R+3a}{2(R+a)}\alpha(0)\right.
\nonumber \\
&&
\phantom{aaa}
+\sum\limits_{l=1}^{\infty}\alpha(i\xi_l)
\int_{\zeta_l}^{\infty}dy\,ye^{-y}\left[y-
\frac{a}{2(R+a)}\right]
\label{eq19} \\
&&
\phantom{aaaaa}
\times\left.
\vphantom{\frac{4R+3a}{2(R+a)}}
\left[2r_{{\rm TM},g}^{(1)}(\zeta_l,y)+
\frac{\zeta_l^2}{y^2}\left(r_{{\rm TE},g}^{(1)}(\zeta_l,y)
-r_{{\rm TM},g}^{(1)}(\zeta_l,y)\right)\right]\right\},
\nonumber \\
&&
C_F(a,T)=\frac{k_BT}{8}\sqrt{\frac{R}{R+a}}\left\{
\frac{3(2R+3a)(2R+a)}{2(R+a)^2}\alpha(0)\right.
\nonumber \\
&&
\phantom{aaa}
+\sum\limits_{l=1}^{\infty}\alpha(i\xi_l)
\int_{\zeta_l}^{\infty}dy\,ye^{-y}\left[y^2-
\frac{3a^2}{4(R+a)^2}\right]
\label{eq20} \\
&&
\phantom{aaaaa}
\times\left.
\vphantom{\frac{3(2R+3a)(2R+a)}{2(R+a)^2}}
\left[2r_{{\rm TM},g}^{(1)}(\zeta_l,y)+
\frac{\zeta_l^2}{y^2}\left(r_{{\rm TE},g}^{(1)}(\zeta_l,y)
-r_{{\rm TM},g}^{(1)}(\zeta_l,y)\right)\right]\right\}.
\nonumber
\end{eqnarray}
\noindent
In terms of new variables (\ref{eq17}) the reflection
coefficients (\ref{eq9}) take the form
\begin{eqnarray}
&&
r_{{\rm TM},g}^{(1)}(\zeta_l,y)=
\frac{2yaK}{2yaK+\zeta_l^2},
\label{eq21} \\
&&
r_{{\rm TE},g}^{(1)}(\zeta_l,y)\equiv
r_{{\rm TE},g}^{(1)}(y)=
\frac{2aK}{2aK+y}.
\nonumber
\end{eqnarray}

To perform computations using Eqs.~(\ref{eq19}) and
(\ref{eq20}) one needs some expressions for the atomic
and molecular dynamic polarizabilities of hydrogen.
As was shown in Refs.~\cite{25,40}, for the calculation of
van der Waals interaction the polarizabilities can be
represented with sufficient precision in the framework
of the single-oscillator model,
\begin{eqnarray}
&&
\alpha(i\xi_l)=\alpha_a(i\xi_l)=
\frac{g_a}{\omega_a^2+\xi_l^2},
\label{eq22} \\
&&
\alpha(i\xi_l)=\alpha_m(i\xi_l)=
\frac{g_m}{\omega_m^2+\xi_l^2},
\nonumber
\end{eqnarray}
\noindent
for a hydrogen atom and a molecule, respectively. Here,
$g_a=\alpha_a(0)\omega_a^2$ with the static atomic
polarizability of hydrogen equal to $\alpha_a(0)=4.50\,$a.u.
and the characteristic frequency $\omega_a=11.65\,$eV
\cite{41}. For a hydrogen molecule it holds
$g_m=\alpha_m(0)\omega_m^2$ with $\alpha_m(0)=5.439\,$a.u.
and $\omega_m=14.09\,$eV \cite{41}. Note that when
substituting polarizabilities in Eqs.~(\ref{eq19}),
(\ref{eq20}), they should be expressed in cubic meters
(1\,a.u. of polarizability is equal to
$1.482\times 10^{-31}\,\mbox{m}^{3}$).

The computations were performed for single-wall carbon
nanotubes with radia $R=2,\,3,\,4$ and 5\,nm at
$T=300\,$K. As an example, in Fig.~2 we plot the van der
Waals coefficient (\ref{eq19}) as a function of separation
in the region from 1 to 3\,nm for the nanotube with
$R=5\,$nm. Solid line 1 labels a nanotube interaction with
a hydrogen atom and solid line 2 with a hydrogen molecule.
For comparison, in the same figure the previously computed
\cite{28} van der Waals coefficients for the interaction of
a hydrogen atom (the dashed line 1) and molecule (the dashed
line 2) with a plane graphene sheet are included.
As is seen in Fig.~2 (the solid lines 1 and 2), at all
separations the van der Waals coefficient of a
molecule-nanotube interaction is larger than of an
atom-nanotube interaction. At the same time, the van der
Waals coefficients for the interaction of a hydrogen atom or
a molecule with graphene (the dashed lines 1 and 2, respectively)
are larger than the respective coefficients for the
interaction with a single-wall carbon nanotube. This excess is
of about 30\%. Note that the obtained results practically do
not depend on temperature in the temperature region from 0
to 300\,K. 

In Table~I we present a few computational results for
the van der Waals coefficient $C_3$ in the case of hydrogen
atom (columns 2--5) and molecule (columns 6--9) interacting
with single-wall carbon nanotubes of different radia.
From Table~I it is seen that the van der Waals coefficient
$C_3$ is monotonously increasing with the increase of
nanotube radius.
As in Fig.~2, the separation distance between an atom
(molecule) and a nanotube varies from 1 to 3\,nm (recall that
at shorter separations some additional forces of chemical
and exchange nature should be taken into account; at larger
separations the accuracy of the obtained theoretical
expressions may decrease).

In Fig.~3 we plot the coefficient $C_F$ for the van der Waals
force defined in Eq.~(\ref{eq20}) 
acting between a hydrogen atom (the
solid line 1) or a molecule (the solid line 2) interacting
with the single-wall carbon nanotube of $R=5\,$nm radius. As it
holds for the coefficient $C_3$, the values of the coefficient
$C_F$ for a molecule are larger than for an atom at all
separations under consideration. Dotted lines in Fig.~3
(labeled 1 for an atom and 2 for a molecule) represent the
results obtained on the assumption that $C_F=3C_3$, i.e.,
on the assumption that the van der Waals coefficient
$C_3=$const and does not depend on separation distance.
As is seen in Fig.~3, the differences between solid and dotted
lines are of about 15--20\%. Thus, the dependence of the van der
Waals coefficients on separation
in atom (molecule)---nanotube interaction
is essential for obtaining the computational results of
high precision.

In Table~II the computational results for the van der Waals force
coefficient $C_F$ are presented. Columns 2--5 are related to the
case of hydrogen atom and columns 6--9 to hydrogen molecule
interacting with single-wall carbon nanotubes of radia
$R=2,\,3,\,4$ and 5\,nm. As in Table~I, the separation
distance varies from 1 to 3\,nm. From Table~II it is seen
that the magnitudes of the coefficient $C_F$ increase with
the increase of a nanotube radius and decrease with the
increase of separation distance. The respective magnitudes of
$C_F$ for a hydrogen molecule are larger than for an atom for 
nanotubes of different radia.

\section{Comparison of interactions between hydrogen atom (molecule)
and carbon nanotube
in the cases of single- and multi-wall 
nanotubes}

In this section we compare the van der Waals coefficients for the
interaction of hydrogen atom or molecule with single-wall and 
multi-wall carbon nanotubes. This permits us to determine how thick 
should be a multi-wall nanotube in order the idealization of
the dielectric permittivities of graphite be applicable.
We will also quantitatively compare the cases of multi-wall
and single-wall carbon nanotubes depending on the separation
distance between a nanotube and a microparticle.

The van der Waals coefficient of the interaction between
a single-wall nanotube and a microparticle is given by
Eq.~(\ref{eq19}) with reflection coefficients (\ref{eq21}).
To obtain the van der Waals coefficient of the interaction
with a multi-wall nanotube, one should replace the
reflection coefficients $r_{{\rm TM,TE},g}^{(1)}$ in Eq.~(\ref{eq19})
with the coefficients $r_{{\rm TM,TE},d}^{(1)}$ defined
in Eq.~(\ref{eq5}) \cite{25}. In terms of dimensionless variables
(\ref{eq17}) the latter take the form
\begin{eqnarray}
&&
r_{{\rm TM},d}^{(1)}(\zeta_l,y)=
\frac{\varepsilon_{xl}\varepsilon_{zl}y^2-
f_z^2(y,\zeta_l)}{\varepsilon_{xl}\varepsilon_{zl}y^2+
f_z^2(y,\zeta_l)+2\sqrt{\varepsilon_{xl}\varepsilon_{zl}}
yf_z(y,\zeta_l)\coth\left[f_z(y,\zeta_l)d/(2a)\right]},
\nonumber \\
&&
r_{{\rm TE},d}^{(1)}(\zeta_l,y)=
\frac{f_x^2(y,\zeta_l)-y^2}{y^2+
f_x^2(y,\zeta_l)+2
yf_x(y,\zeta_l)\coth\left[f_x(y,\zeta_l)d/(2a)\right]},
\label{eq23}
\end{eqnarray}
\noindent
where the following notations are introduced
\begin{eqnarray}
&&
\varepsilon_{xl}\equiv\varepsilon_x(i\zeta_l\omega_c),
\qquad
\varepsilon_{zl}\equiv\varepsilon_z(i\zeta_l\omega_c),
\label{eq24} \\
&&
f_z^2(y,\zeta_l)=y^2+\zeta_l^2(\varepsilon_{zl}-1),
\quad
f_x^2(y,\zeta_l)=y^2+\zeta_l^2(\varepsilon_{xl}-1).
\nonumber
\end{eqnarray}

The computational results for
the van der Waals coefficient $C_3$ of atom-nanotube
interaction are presented in Fig.~4 by the solid dots 1,\,2,\,3
(the single-wall carbon nanotube of 5\,nm radius at separations
$a=1,\,2$ and 3\,nm to a hydrogen atom, respectively) and 
by the solid dots connected with solid lines for the
multi-wall carbon nanotubes of 5\,nm external radius with
$n=1,\,2,\,3,\,4$ and 5 walls, respectively.
For a single-wall nanotube computations were performed
using Eqs.~(\ref{eq19}), (\ref{eq21}) and for a multi-wall
nanotubes using Eqs.~(\ref{eq19}), (\ref{eq23}). In both
cases the dynamic atomic polarizability $\alpha_a$ is
taken in Eq.~(\ref{eq22}) and for a multi-wall nanotube
$d=3.4(n-1)\,$\AA. As is seen in Fig.~4, the van der Waals
coefficient computed by the formula
for a multi-wall nanotube with only
one wall does not coincide with the coefficient computed
by the formula for a single-wall nanotube. This is expected
result because the reflection coefficients in Eq.~(\ref{eq23})
approach zero when the nanotube thickness $d$ vanishes.
At the same time, as Fig.~4 suggests, the van der Waals
coefficients for a multi-wall nanotube with $n=3$ walls at
1\,nm from an atom and for a multi-wall nanotube with
$n=2$ walls at a separation 2 or 3\,nm from an atom are
in the proper proportion to the coefficients computed
for a single-wall nanotube using the reflection coefficients
(\ref{eq21}) (the analogical result was obtained in
Ref.~\cite{28} for the interaction of a graphene and a graphite
plate with an Au semispace). This permits us to conclude
that the macroscopic concept of graphite dielectric
permittivity used in Ref.~\cite{25} to describe multi-wall
carbon nanotubes is already applicable for nanotubes
containing only two or three walls depending on separation
distance between a nanotube and an atom.

Fig.~5 contains the same information,  as in Fig.~4, but
for a hydrogen molecule. It is seen that
the approximation of graphite
dielectric permittivities is also well applicable for
multi-wall nanotube with 3 walls spaced
1\,nm apart from a molecule and
for two-wall nanotube at a separation of 2 or 3\,nm from a molecule.
The values of the van der Waals coefficient for a molecule are, 
however, larger than for an atom in accordance to Sec.~III.
The single-wall nanotube cannot be described using the
macroscopic concept of dielectric permittivity but as a
cylindrical plasma sheet with appropriate boundary conditions. 

\section{Conclusions and discussion}

In the foregoing, we have obtained the Lifshitz-type formulas
describing the free energy and force of  
the van der Waals interaction of a hydrogen atom
or a molecule with a single-wall carbon nanotube. The
nanotube is represented
using the description of a cylindrical graphene sheet in
terms of two-dimensional plasma carrying free
$\pi$-electrons with appropriate boundary conditions imposed
on the electromagnetic field. Recently this approach was used
to describe the interaction of a flat graphene sheet with a
material plate or a microparticle, and a material plate with
a single-wall carbon nanotube \cite{28}. The formulas obtained
here were applied to calculate the van der Waals free energy
and force in the configuration of a hydrogen atom or a molecule
interacting with single-wall carbon nanotubes of different
radia. The comparison with the previously developed
Lifshitz-type formalism applicable to multi-wall carbon
nanotubes \cite{25} have allowed to determine the application limits
of the macroscopic concept of graphite dielectric permittivity.
The latter was shown to be applicable even to two- or
three-wall nanotubes depending on the separation distance
between a nanotube and an atom (molecule).

The developed formalism suggests some advantages in comparison
to phenomenological and semi-phenomenological approaches used
to calculate the van der Waals interaction.  As was
recently noted in Ref.~\cite{11}, the phenomenological method
of pairwise summation of the intermolecular van der Waals
potentials can lead to even qualitatively wrong results for
the nonretarded van der Waals interaction between metallic
nanotubes and nanowires (in Ref.~\cite{45} it was
emphasized that the pairwise summation of intermolecular
potentials leads to correct dependence of the van der Waals
potential and force on separation distance only for bulk 
three-dimensional bodies).

The possibility to use the Lifshitz theory of dispersion
forces for the investigation of the van der Waals interaction 
between hydrogen atoms (molecules) and single-wall carbon
nanotubes demonstrated in the paper may be helpful
for the problem of hydrogen storage. According to recent
review on the subject \cite{46}, there are conceptual
possibilities to create carbon nanostructures capable of
absorbing more than 10 mass \% of hydrogen. The resolution
of this fundamental problem requires a detailed
investigation of the microscopic mechanisms of the
interaction between hydrogen and graphite.
To attain these ends the above approach should be
extended to distances below 1\,nm. At so small separations 
there may be attractive chemical forces and short-range
repulsive forces of exchange nature. These forces depend
on atomic structure and cannot be taken into account
by means of the boundary conditions. Some approximate
description is possible by combining the energy of
van der Waals attraction with phenomenological
potentials \cite{47a}.

\section*{Acknowledgments}
This work was supported by
the Russian Foundation for
Basic Research (Grant No. 05--08--18119a).
G.L.K. and V.M.M. were also partially supported by
Deutsche Forschungsgemeinschaft
(Grant No. 436\,RUS\,113/789/0--3).

\begin{figure*}
\vspace*{-7cm}
\includegraphics{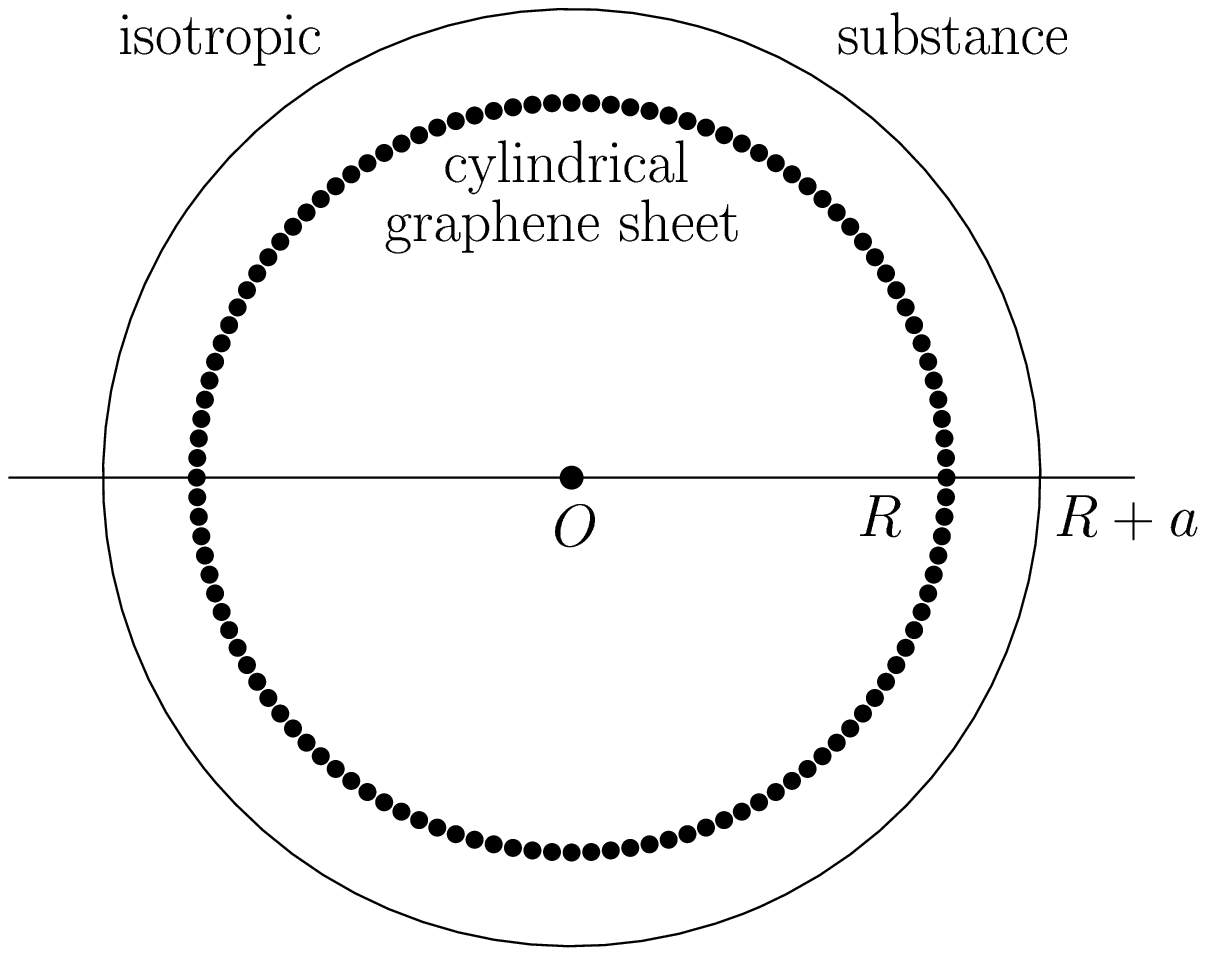}
\vspace*{-9cm} \caption{Schematic of the cylindical graphene sheet
of radius $R$ which is concentrically placed into a cylindrical
cavity of radius $R+a$ in the infinite space filled with an
isotropic substance.
}
\end{figure*}

\begin{figure*}
\vspace*{-7cm}
\includegraphics{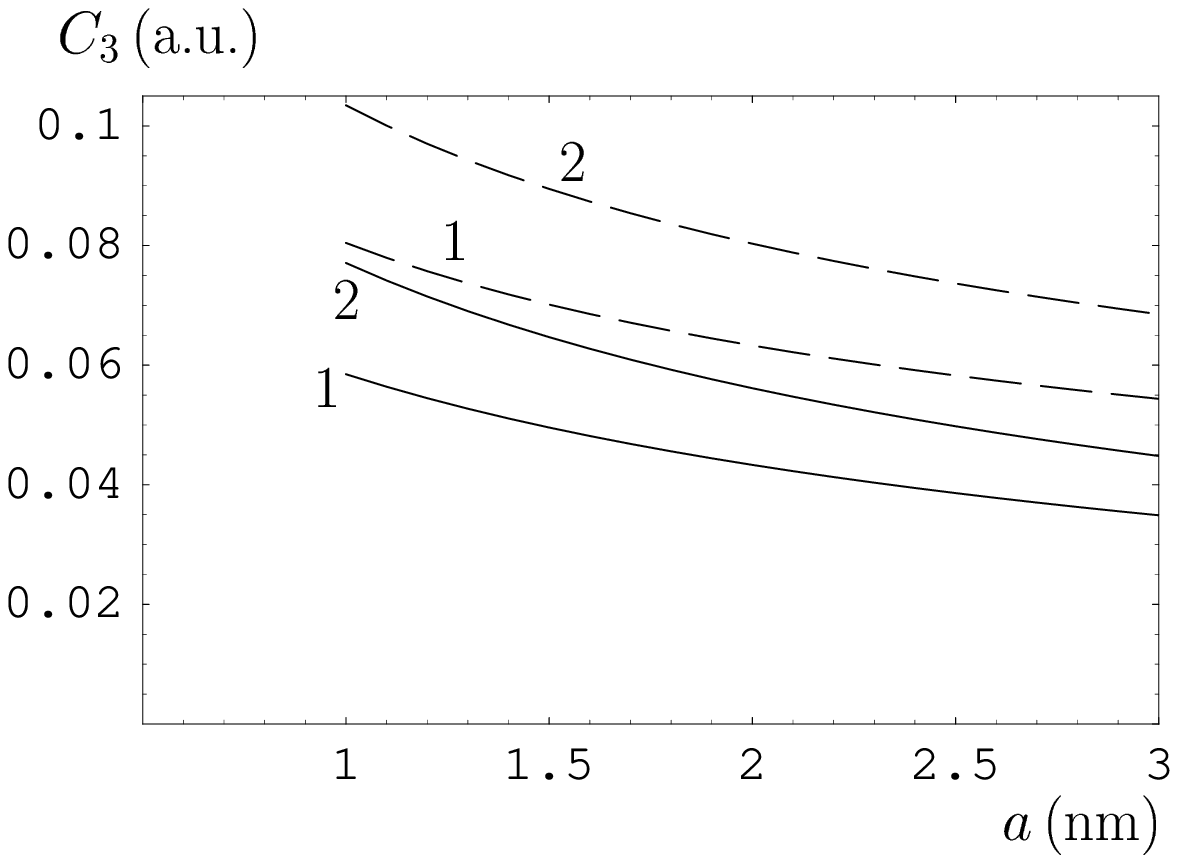}
\vspace*{-10cm} 
\caption{The  van der Waals coefficient as a function of separation 
for the interaction of a hydrogen atom (lines labeled 1)
or a molecule (lines labeled 2) with the
single-wall carbon nanotube of $R=5\,$nm radius (solid lines)
and with a plane graphene sheet (dashed lines).}
\end{figure*}

\begin{figure*}
\vspace*{-7cm}
\includegraphics{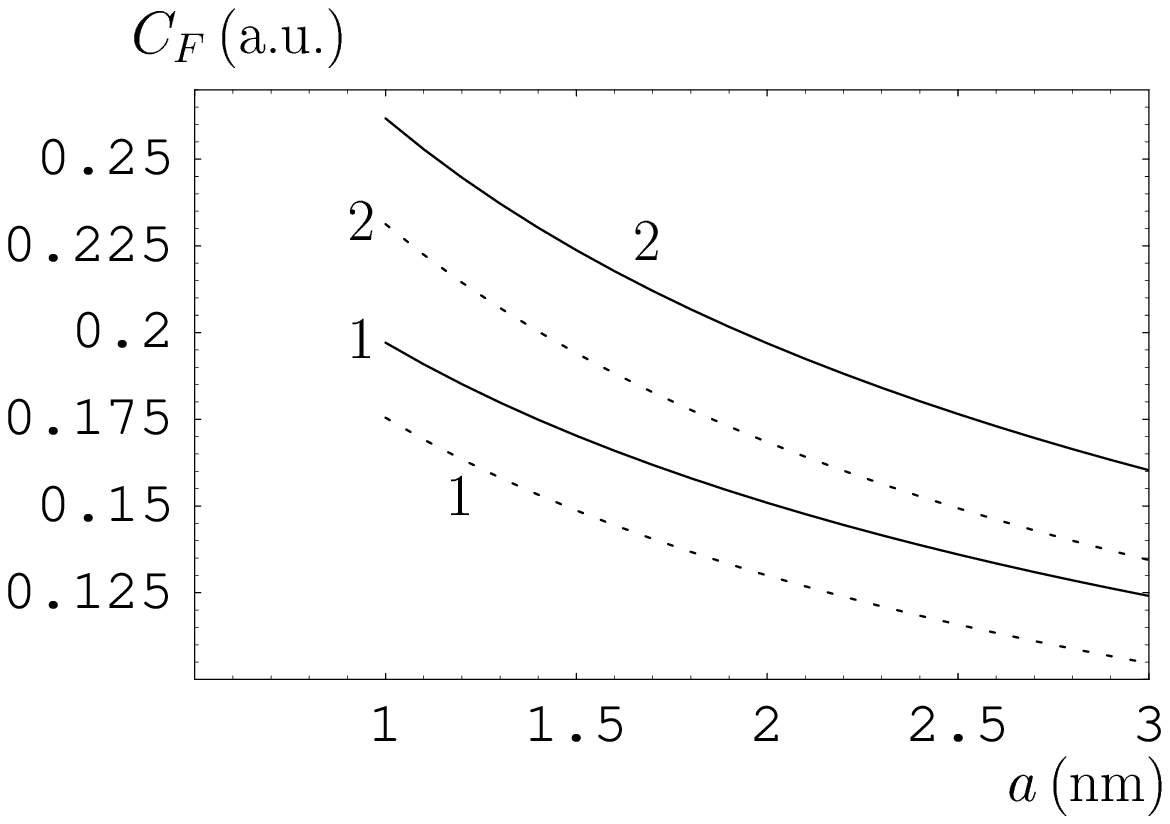}
\vspace*{-10cm} 
\caption{The coefficient $C_F$ as a function of separation 
for the interaction of a hydrogen atom (lines labeled 1)
or a molecule (lines labeled 2) with the
single-wall carbon nanotube of $R=5\,$nm radius (solid lines).
Dotted lines are drown under the assumption that
$C_F=3C_3$, i.e., that the van der Waals coefficient $C_3$
does not depend on separation.}
\end{figure*}

\begin{figure*}
\vspace*{-7cm}
\includegraphics{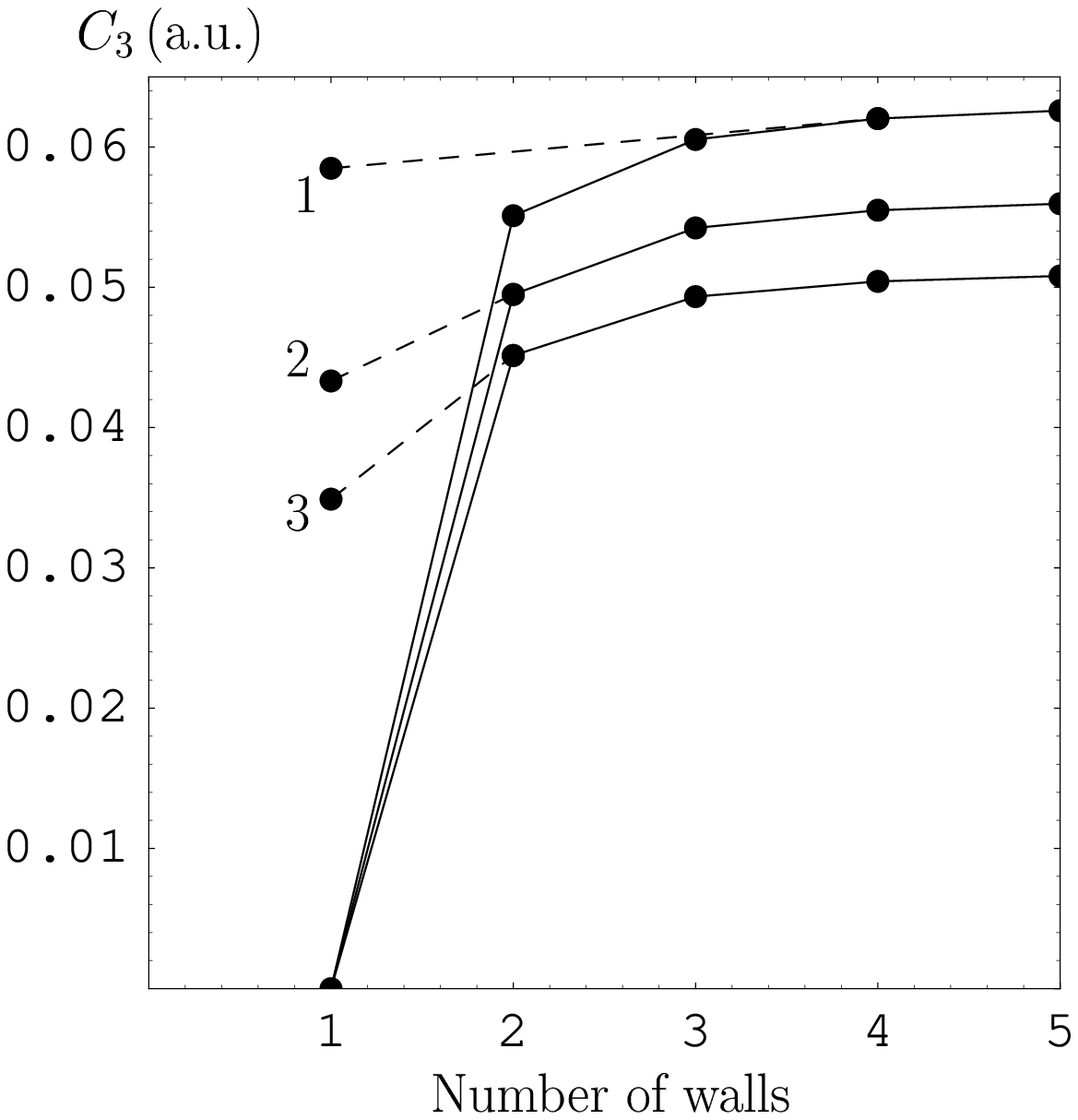}
\vspace*{-9cm} \caption{The van der Waals coefficient 
as a function of the number of walls for the interaction of 
a hydrogen atom with the multi-wall carbon nanotube of
$R=5\,$nm external radius (solid dots connected with
solid lines) and with a single-wall carbon nanotube of the
same radius (solid dots 1,\,2,\,3) spaced at 1,\,2 and 3\,nm
from the atom, respectively.}
\end{figure*}

\begin{figure*}
\vspace*{-7cm}
\includegraphics{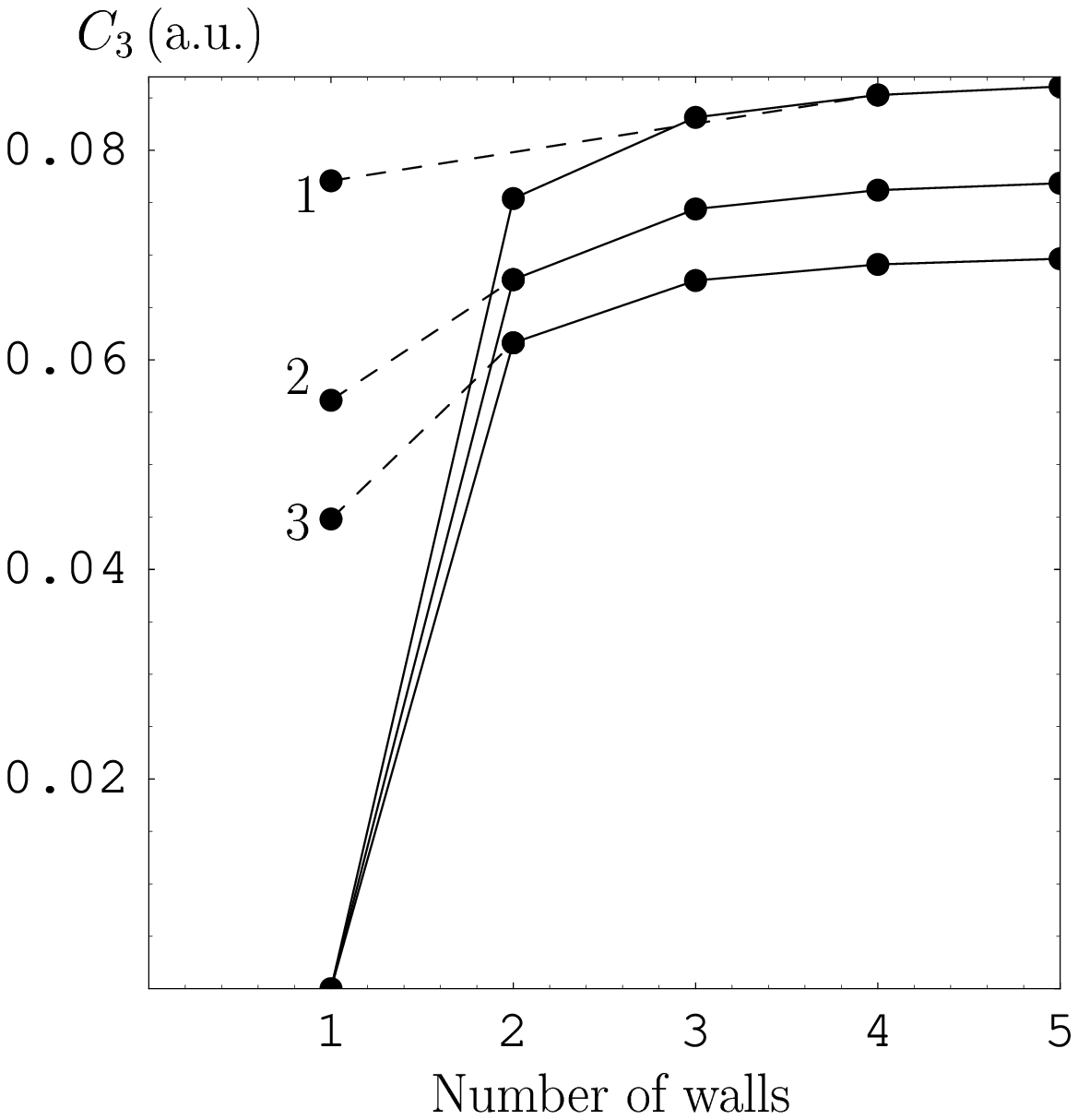}
\vspace*{-9cm} \caption{The van der Waals coefficient 
as a function of the number of walls for the interaction of 
a hydrogen molecule with the multi-wall carbon nanotube of
$R=5\,$nm external radius (solid dots connected with
solid lines) and with a single-wall carbon nanotube of the
same radius (solid dots 1,\,2,\,3) spaced at 1,\,2 and 3\,nm
from the molecule, respectively.}
\end{figure*}
\begingroup
\squeezetable
\begin{table}
\caption{The van der Waals coefficient as a function of separation 
for the interaction of a hydrogen atom or a molecule with
single-wall carbon nanotubes of different radia.}
\begin{ruledtabular}
\begin{tabular}{ccccccccc}
&\multicolumn{8}{c}{$C_3\,$(a.u.)}\\
\cline{2-9}
$a$&\multicolumn{4}{c}{hydrogen atom}&
\multicolumn{4}{c}{hydrogen molecule}\\
\cline{2-9}
(nm)&$R=2\,$nm&$R=3\,$nm&$R=4\,$nm&$R=5\,$nm
&$R=2\,$nm&$R=3\,$nm&$R=4\,$nm&$R=5\,$nm\\
\hline
1.0&0.0503&0.0544&0.0569&0.0585&0.0664&0.0718&0.0750&0.0771\\
1.2&0.0460&0.0502&0.0528&0.0545&0.0604&0.0659&0.0692&0.0715\\
1.4&0.0424&0.0466&0.0493&0.0511&0.0554&0.0610&0.0644&0.0668\\
1.6&0.0393&0.0436&0.0463&0.0482&0.0513&0.0568&0.0603&0.0627\\
1.8&0.0367&0.0410&0.0437&0.0456&0.0478&0.0532&0.0568&0.0592\\
2.0&0.0345&0.0387&0.0414&0.0433&0.0477&0.0501&0.0536&0.0561\\
2.2&0.0325&0.0366&0.0394&0.0413&0.0420&0.0474&0.0509&0.0534\\
2.4&0.0307&0.0348&0.0375&0.0395&0.0397&0.0449&0.0484&0.0509\\
2.6&0.0292&0.0332&0.0358&0.0378&0.0376&0.0427&0.0462&0.0487\\
2.8&0.0277&0.0317&0.0343&0.0363&0.0357&0.0407&0.0442&0.0467\\
3.0&0.0265&0.0303&0.0330&0.0349&0.0340&0.0389&0.0423&0.0448
\end{tabular}
\end{ruledtabular}
\end{table}
\endgroup
\begingroup
\squeezetable
\begin{table}
\caption{The coefficient $C_F$ as a function of separation 
for the van der Waals force acting between a hydrogen atom or 
a molecule and
single-wall carbon nanotubes of different radia.}
\begin{ruledtabular}
\begin{tabular}{ccccccccc}
&\multicolumn{8}{c}{$C_F\,$(a.u.)}\\
\cline{2-9}
$a$&\multicolumn{4}{c}{hydrogen atom}&
\multicolumn{4}{c}{hydrogen molecule}\\
\cline{2-9}
(nm)&$R=2\,$nm&$R=3\,$nm&$R=4\,$nm&$R=5\,$nm
&$R=2\,$nm&$R=3\,$nm&$R=4\,$nm&$R=5\,$nm\\
\hline
1.0&0.175&0.186&0.193&0.197&0.232&0.248&0.256&0.262\\
1.2&0.162&0.174&0.181&0.185&0.214&0.230&0.239&0.245\\
1.4&0.150&0.163&0.170&0.175&0.198&0.214&0.224&0.231\\
1.6&0.140&0.153&0.161&0.166&0.184&0.201&0.211&0.218\\
1.8&0.132&0.145&0.153&0.158&0.173&0.190&0.200&0.207\\
2.0&0.124&0.138&0.146&0.151&0.162&0.180&0.190&0.197\\
2.2&0.118&0.131&0.139&0.144&0.154&0.170&0.181&0.189\\
2.4&0.112&0.125&0.133&0.139&0.146&0.162&0.173&0.181\\
2.6&0.108&0.120&0.128&0.133&0.138&0.155&0.166&0.174\\
2.8&0.102&0.115&0.123&0.128&0.132&0.148&0.159&0.167\\
3.0&0.0975&0.110&0.119&0.124&0.126&0.142&0.153&0.161
\end{tabular}
\end{ruledtabular}
\end{table}
\endgroup
\end{document}